\documentclass[pra,secnumarabic,groupedaddress,showpacs]{revtex4}
\usepackage{amsmath}
\usepackage[latin1]{inputenc}
\newcommand{\vect}[1]{\mathbf{#1}}
\newcommand{\gvect}[1]{\boldsymbol{#1}}
\begin{document}

\title{A Refutation of Bell's Theorem\footnote[1]{This work was
presented at the international conference "Foundations of
Probability and Physics" held in Växjö, Sweden; Nov. 2000.}}%
\author{Guillaume ADENIER}
\email{guillaume.adenier@ulp.u-strasbg.fr} \affiliation{Dept. of
Physics, Louis Pasteur University, Strasbourg, France.}
\date{January 18, 2001}

\begin{abstract} Bell's Theorem was developed on the basis of
considerations  involving a linear combination of spin correlation
functions, each of which has a distinct pair of arguments. The
simultaneous presence of these different pairs of arguments in the
same equation can be understood in two radically different ways:
either as `strongly objective,' that is, all correlation functions
pertain to the same set of particle pairs, or as `weakly
objective,' that is, each correlation function pertains to a
different set of particle pairs.

It is demonstrated that once this meaning is determined, no
discrepancy appears between local realistic theories and quantum
mechanics: the discrepancy in Bell's Theorem is due only to a
meaningless comparison between a local realistic inequality
written within the strongly objective interpretation (thus
relevant to a single set of particle pairs) and a quantum
mechanical prediction derived from a weakly objective
interpretation (thus relevant to several different sets of
particle pairs). \end{abstract} \maketitle


\section{Introduction}

Bell's Theorem\cite{JSB2} exhibits a peculiar discrepancy between
any local realistic theory and Quantum Mechanics, which leads to
empirically distinguishable alternatives. The quandary is that
neither local realistic  conceptions nor Quantum Mechanics are
easy to abandon. Indeed, classical physics and common sense are
usually based upon the former, while the latter is rightly
presented as the most successful theory of all times. Several
experiments have been done, all but a few\cite{FS1} show
violations of Bell inequalities. Yet, the ideas brought forth by
Bell's Theorem are so disconcerting that there is still
incredulity, not to mention antipathy, evoked by the verdict. The
purpose of this article is to provide a refutation of this
theorem, within a strictly quantum theoretical framework, without
the use of outside assumptions.

Although experiments showing violations of Bell's inequalities are
getting increasingly accurate and loophole-free\cite{ASP1}, it
must be stressed that they, no matter how accurate and close to
the ideal, can prove no more than the  validity of quantum
mechanics, and not the validity of the theorem. Herein, it will be
assumed that all tests conducted so far prove conclusively the
validity of Quantum Mechanics. In other words, the purpose of this
article is not to criticise the numerous experiments, or quantum
mechanics for that matter, but Bell's Theorem itself.

\section{The EPRB gedanken experiment}\label{EPRB}

\subsection{Spin observables and the singlet state}\label{observing}

Bell's theorem is usually based on a didactic reformulation of the
EPR (Einstein, Podolsky and Rosen\cite{EPR1}) gedanken experiment,
due to D. Bohm\cite{DB1}.  In this EPRB gedanken experiment, a
pair of spin-½ particles with total spin zero is produced such
that each particle moves away from the source  in opposite
directions along the y-axis. Two Stern-Gerlach devices are placed
at opposite points (left and right) on the y-axis, and are
oriented respectively along the directions $\vect{u}$ and
$\vect{v}$.

The spin observable associated with a measurement given by a
Stern-Gerlach device oriented along the unit vector $\gvect{u}$ is
$\gvect{\sigma}\cdot\vect{u}$ (to simplify notation, ${\hbar}/{2}$
is set to $1$ throughout); where the components of
$\gvect{\sigma}$ are then the Pauli matrices $\sigma_x$,
$\sigma_y$, and $\sigma_z$. Let $\mathcal{H}_\mathrm{L}$ and
$\mathcal{H}_\mathrm{R}$ be the Hilbert spaces associated with
each Stern-Gerlach device respectively. The Hilbert space
$\mathcal{H}$ associated with the entire EPRB system is the direct
product of the Hilbert spaces associated with each Stern-Gerlach
device:
\begin{equation}\label{ep}
    \mathcal{H}\equiv\mathcal{H}_\mathrm{L}\otimes\mathcal{H}_\mathrm{R}.
\end{equation}
The spin observables relevant to $\mathcal{H}_\mathrm{L}$ and
$\mathcal{H}_\mathrm{R}$ have their respective counterpart in this
new product space $\mathcal{H}$ as
\begin{subequations}
    \label{prols}
    \begin{eqnarray}
\gvect{\sigma}_\mathrm{L}\cdot\vect{u}\equiv\gvect{\sigma}\cdot\vect{u}\otimes
        1\negmedspace \mathrm{l}_\mathrm{R},\label{s1}\\
        \gvect{\sigma}_\mathrm{R}\cdot\vect{v}\equiv
        1\negmedspace
        \mathrm{l}_\mathrm{L}\otimes\gvect{\sigma}\cdot\vect{v},\label{s2}
    \end{eqnarray}
\end{subequations}
where $1\negmedspace \mathrm{l}_\mathrm{L}$ and $1\negmedspace
\mathrm{l}_\mathrm{R}$ are the identity operators of
$\mathcal{H}_\mathrm{L}$ and $\mathcal{H}_\mathrm{R}$. Contrary to
the observables $\gvect{\sigma}\cdot\vect{u}$ and
$\gvect{\sigma}\cdot\vect{v}$ which are mutually non commuting
when $\vect{u}\neq\vect{v}$, these new observables
$\gvect{\sigma}_\mathrm{L}\cdot\vect{u}$ and
$\gvect{\sigma}_\mathrm{R}\cdot\vect{v}$ do commute, reflecting
the fact that the Stern-Gerlach devices are arbitrarily far from
each other, and are thus measuring distinct subsystems. The
product of these two observables
\begin{equation}\label{spincor}
(\gvect{\sigma}_\mathrm{L}\cdot\vect{u})(\gvect{\sigma}_\mathrm{R}\cdot\vect{v})=
\gvect{\sigma}\cdot\vect{u}\otimes\gvect{\sigma}\cdot\vect{v}
\end{equation}
is therefore also an observable and can be understood as a
\emph{spin correlation observable} corresponding to the
\emph{joint spin measurement} of both Stern-Gerlach devices.

The product space $\mathcal{H}$ is spanned by the product basis
formed by the four eigenvectors $\{ |++\rangle, |+-\rangle,
|-+\rangle, |--\rangle \}$ associated with the spin correlation
observable
$(\gvect{\sigma}_\mathrm{L}\cdot\vect{n})(\gvect{\sigma}_\mathrm{R}\cdot\vect{n})$
where $\vect{n}$ is a unitary vector. In an EPRB gedanken
experiment, the source produces particle pairs with zero total
spin, represented by the singlet state
\begin{equation}\label{enf}
|\psi\rangle=\frac{1}{\sqrt{2}}\Big[|+-\rangle-|-+\rangle\Big].
\end{equation} This singlet state has the important property of
being invariant under rotation, which permits one to ignore the
explicit form  of $\vect{n}$ in expressing the $\mathcal{H}$ basis
(see, for instance, Ref.  \cite{GHSZ1}).

\subsection{Statistical properties of the singlet state}

As it is, nothing certain can be said either about a single spin
measurement, or about a single spin correlation measurement,
performed on a system represented by the singlet state. According
to the Born interpretation of the state vector, only probabilistic
predictions---such as expectation values relevant to numerous
measurements in the same context---are allowed.

It can be shown (see, for instance, Ref. \cite{ctdl1}, chapter
IV), that the expectation value of an observable $\hat{A}$ is
${\langle \hat{A} \rangle}_{\phi}=\langle\phi | \hat{A} |
\phi\rangle$ and therefore, with the help of Eqs. (\ref{prols})
and (\ref{enf}), that the \emph{expectation value of a spin
observable} for the singlet state $|\psi\rangle$ is zero:
\begin{subequations}
    \label{expspin}
 \begin{eqnarray}
  \langle\gvect{\sigma}_\mathrm{L}\cdot\vect{u}\rangle_\psi &=&
  \langle\psi|\gvect{\sigma}\cdot\vect{u}\otimes1\negmedspace\mathrm{l}_\mathrm{R}|\psi\rangle=0,
  \\
  \langle\gvect{\sigma}_\mathrm{R}\cdot\vect{v}\rangle_\psi &=&
  \langle\psi|1\negmedspace\mathrm{l}_\mathrm{L}\otimes\gvect{\sigma}\cdot\vect{v}|\psi\rangle=0,
 \end{eqnarray}
\end{subequations}
whatever $\vect{u}$ and $\vect{v}$, as follows from the rotational
invariance of the singlet state. Likewise, the \emph{expectation
value of the spin correlation observable} is
\begin{subequations}
    \label{qcor}
 \begin{eqnarray}
    \label{qcor1}
    E^\psi(\vect{u},\vect{v})=&
    \langle\psi|
    (\gvect{\sigma}_\mathrm{L}\cdot\vect{u})
    (\gvect{\sigma}_\mathrm{R}\cdot\vect{v})
    |\psi\rangle
    \\                  \label{qcor2}
    =&-\vect{u}\cdot\vect{v},
 \end{eqnarray}
\end{subequations}
which depends only on the relative angle between  $\vect{u}$ and
$\vect{v}$ (see, for instance, Refs. \cite{FS1}, \cite{GHSZ1}, or
\cite{AB1}).

\subsection{Perfect correlations and
hidden-variables}\label{perfectc}

When $\vect{u}=\vect{v}$, the expectation value of the spin
correlation observable (\ref{qcor}) is equal to $-1$, meaning that
if both Stern-Gerlach devices are oriented along the same
direction, then with certainty the outcomes will be found to be
opposite. Since the Stern-Gerlach devices are arbitrarily far from
each other, a perfect correlation can be understood from a
realistic point of view as invested in the particles at their
inception.  This, however, would mean that the singlet state is
incomplete, and therefore, that it should be possible to give a
more precise specification using additional ``hidden-variables''.
On the other hand, if a more complete description is impossible,
then this perfect correlation seems rather mysterious, since a
measurement performed on one of the subsystems seems to be capable
of influencing the measurement on the other subsystem, whatever
the distance between them.

In order to facilitate a choice between incompleteness and
non-locality of Quantum Mechanics, Bell's idea was to specify
mathematical requirements for a generic local hidden-variables
theory, and then to compare its predictions with those from
quantum mechanics and the results of experiments.

In a local realistic hidden-variables model, a single particle
pair is thus supposed to be entirely characterised by means of a
set of hidden-variables, which are symbolically represented by a
parameter $\lambda$, so that the measurement result on the left
along $\vect{u}$ can be written as $A(\vect{u},\lambda)$, and the
result on the right along $\vect{v}$ as $B(\vect{v},\lambda)$.
Although the hidden-variables model is supposed to be fully
deterministic, it must also be capable of reproducing the
stochastic nature of the EPRB gedanken experiment expressed in
Eqs.  (\ref{expspin}) and (\ref{qcor}). For that purpose, the
complete state specification $\lambda_i$ of any particle pair with
label $i$ must be a random variable: its complete state
$\lambda_i$ is supposed to be drawn randomly according to a
probability distribution $\rho$ (see Refs. \cite{JSB2} and
\cite{JSB4}), meaning that the probability of having $\lambda_i$
equal to a particular $\lambda$ is $\rho (\lambda)$.

Consider a set of $N$ particle pairs $\{i=1,\ldots,N\}$, the
\emph{mean value of joint spin measurements} for this set is :
\begin{equation}\label{objcor2}
    M^\rho(\vect{u},\vect{v})=
    \frac{1}{N}\sum_{i=1}^{N}
    A(\vect{u},\lambda_i)B(\vect{v},\lambda_i).
\end{equation}
The probability distribution $\rho$ is supposed to assure the
equality between this mean value and the expected value, Eq.
(\ref{qcor2}), given by Quantum Mechanics when $N$ goes to
infinity.

\section{The `CHSH' function}\label{CHSH}

In order to establish Bell's Theorem, a linear combination of
correlation functions $c(\vect{a},\vect{b})$ with \emph{different
arguments}\cite{JSB8} is considered, once when these correlation
functions are expectation values $E^\psi(\vect{u},\vect{v})$ given
by Quantum Mechanics; i.e., Eq. (\ref{qcor}), and once when they
are mean values $M^\rho(\vect{u},\vect{v})$ given by local
hidden-variables theories, Eq. (\ref{objcor2}); then the results
are to be compared. A well known choice of such a linear
combination is the CHSH (Clauser, Horne, Shimony and Holt
\cite{CHSH1}) function, written with four pairs of arguments:

\begin{equation}\label{CHSHeq}
  S\equiv|
   c(\vect{a},\vect{b})
  -c(\vect{a},\vect{b'})
  +c(\vect{a'},\vect{b})
  +c(\vect{a'},\vect{b'})|.
\end{equation}

Bell's Theorem consists in showing that the quantum prediction for
the CHSH function violates the maximum possible value given by any
local realistic hidden-variables theory. Thus, no such theory will
ever be capable of explaining or reproducing these quantum
results. Herein, this claim is refuted by demanding the rules of
Quantum Mechanics be consistently and meaningfully applied.

To begin, the exact meaning of the simultaneous presence of
\emph{different arguments} in a CHSH function must be clarified.
Basically, there are two possible interpretations, the
\emph{strongly objective} interpretation and the \emph{weakly
objective} interpretation \cite{BDE2,BDE1}:

\begin{description}

\item[Strongly Objective Interpretation] implies that all correlation functions are relevant to the same set of $N$ particle pairs, that is, all four
pairs of directions are considered  simultaneously relevant to
each particle pair. As such they cannot be relevant to  actual
experiments but rather with what result \emph{would have been}
obtained if measured on the same set of $N$ particle pairs along
different directions.

\item[Weakly Objective Interpretation] implies that each
correlation function is actually  to be measured on  distinct sets
of $N$ particle pairs. Each set of $N$ particle pairs pertains to
only one pair of arguments, that is, for each pair only one joint
spin measurement is executed.
\end{description}

The CHSH function was developed specifically for experimental
convenience\cite{CHSH1}. Many experiments  have been done (the
most famous being Aspect's \cite{ADR1}) obviously invoking the
natural interpretation,  namely  the weakly objective one.
Nevertheless, the strongly objective interpretation must also be
considered, since it remains a possible interpretation \emph{a
priori}, and since the choice between strong and weak objectivity
is not made at all explicit in many papers,  including Bell's.

It must be stressed, moreover, that these interpretations are
radically different, not only epistemologically, but also
physically. Indeed, the strongly objective interpretation pertains
to  a single set of $N$ particle pairs characterised by the
corresponding set of parameters $\{\lambda_i\; ; \; i=1,\ldots,N
\}$; whereas the weakly objective interpretation pertains to no
less than 4 sets of $N$ particle pairs. The fact is that a finite
set of $N$ particle pairs characterised by $\{\lambda_i\}$ can't
be identically reproduced, either theoretically (for each complete
state $\lambda_i$ of any particle pair $i$ is a random variable,
as defined in Section \ref{perfectc}), or empirically (for the
experimenter has no control over the complete state of a particle
pair in a singlet state). Of course, when $N$ goes to infinity,
these four sets of $N$ particle pairs necessarily converge to the
same ideal set described by the probability distribution $\rho$.
However, as soon as real experiments are concerned, then
$N\neq\infty$ and these four sets are necessarily \emph{four
different sets of particle pairs} (see Ref.  \cite{AB1}, page 348)
respectively characterised by four different sets of
hidden-variables parameters $\{\lambda_{1,i}\}$,
$\{\lambda_{2,i}\}$, $\{\lambda_{3,i}\}$ and $\{\lambda_{4,i}\}$
(for an alternate approach, see \cite{Khrennikov1}). The
difference between each interpretation can therefore be embodied
in the number of degrees of freedom of the whole system. Let $f$
be the degrees of freedom of a single particle pair. In the
strongly objective interpretation the degrees of freedom of the
whole CHSH system is then $Nf$, whereas in the weakly objective
interpretation it is 4 times as large, that is, $4Nf$.

Thus, before initiating Bell's analysis, one has to choose
explicitly one interpretation and stick to it. Unfortunately, this
is not what has been done. It will be shown here that the
discrepancy exhibited by Bell's Theorem is due to a meaningless
comparison between strongly objective and weakly objective
results, which means comparing the numerical value of the CHSH
function for two systems, one having $Nf$ degrees of freedom, the
other $4Nf$.

\section{The Strongly objective interpretation: counterfactual properties of $N$
particle pairs}\label{obj}

\subsection{A Local realistic inequality within the strongly objective
interpretation}

The local realistic formulation of the CHSH function within strong
objectivity is written
\begin{alignat}{2}
    S^\rho_{\text{strong}}=\Big|
     M^\rho(\vect{a},\vect{b})
    -M^\rho(\vect{a},\vect{b'})
    +M^\rho(\vect{a'},\vect{b})
    +M^\rho(\vect{a'},\vect{b'})
    \Big|,
\end{alignat}
or explicitly (using Eq. \ref{objcor2})
\begin{equation}\label{Srhoobj0}
\begin{split}
    S^\rho_{\text{strong}}=\bigg|\frac{1}{N}\sum_{i=1}^{N}
    &A(\vect{a},\lambda_i)B(\vect{b},\lambda_i)
    -A(\vect{a},\lambda_i)B\vect{b',\lambda_i)}
    \\
    +&A(\vect{a'},\lambda_i)B(\vect{b},\lambda_i)
     +A(\vect{a'},\lambda_i)B(\vect{b'},\lambda_i)\bigg|,
\end{split}
\end{equation}
which after factorisation becomes
\begin{equation}\label{Srhoobj1}
\begin{split}
    S^\rho_{\text{strong}}=
    \bigg|
    \frac{1}{N}\sum_{i=1}^{N}
      &A(\vect{a},\lambda_i)\Big[B(\vect{b},\lambda_i)-B(\vect{b',\lambda_i)}\Big]
      \\
     +&A(\vect{a'},\lambda_i)\Big[B(\vect{b},\lambda_i)+B(\vect{b'},\lambda_i)\Big]
    \bigg|,
\end{split}
\end{equation}
where each term can have two values in the summation
\cite{FS1,AB1}
\begin{equation}\label{rhoobj1}
\begin{split}
     A(\vect{a},\lambda_i)&\Big[B(\vect{b},\lambda_i)-B(\vect{b',\lambda_i)}\Big]
     \\
    +&A(\vect{a'},\lambda_i)\Big[B(\vect{b},\lambda_i)+B(\vect{b'},\lambda_i)\Big]
    =\pm2,
\end{split}
\end{equation}
so that the most restrictive local realistic inequality within the
strongly objective interpretation is :
\begin{equation}\label{locinobj}
    S^{\rho}_{\text{strong}}\leq2.
\end{equation}

This is the well known generalised formulation of Bell's
inequality due to CHSH \cite{CHSH1}. It must be stressed once
more, however, that this inequality has been established only
within the strongly objective interpretation, which means that
each expectation value is relevant to the \emph{same} set of $N$
particle pairs. Hence, this result cannot be compared directly
with results from real experimental tests, where in fact mean
values from four distinct sets of $N$ particle pairs are measured.
The question whether the same inequality can be applied to real
experiments will be discussed in Section \ref{lrweak} (weak
objectivity).

\subsection{The Quantum mechanical prediction within the strongly
objective interpretation}\label{qmobj}

The quantum prediction for the \mbox{CHSH} function within the
strongly objective interpretation is written
\begin{equation}\label{bellq}
  S^{\psi}_{\text{strong}}=|
   E^\psi(\vect{a},\vect{b})
  -E^\psi(\vect{a},\vect{b'})
  +E^\psi(\vect{a'},\vect{b})
  +E^\psi(\vect{a'},\vect{b'})|.
\end{equation}
This equation is usually directly evaluated by replacing each
expectation value by the scalar product result of Eq.
(\ref{qcor2}). This, unfortunately, is all too hasty.

Indeed, in order to understand better the quantum mechanical
meaning of Eq.  (\ref{bellq}), it is advantageous to take a step
backward using Eq.  (\ref{qcor1})
\begin{equation}\label{bof}
 \begin{split}
     S^\psi_{\text{strong}}=
     \Big|
     \langle\psi|
      (\gvect{\sigma}_\mathrm{L}\cdot\vect{a})&(\gvect{\sigma}_\mathrm{R}\cdot\vect{b})
      |\psi\rangle
    -\langle\psi|
     (\gvect{\sigma}_\mathrm{L}\cdot\vect{a})(\gvect{\sigma}_\mathrm{R}\cdot\vect{b'})
     |\psi\rangle
    \\
    +&\langle\psi|
     (\gvect{\sigma}_\mathrm{L}\cdot\vect{a'})(\gvect{\sigma}_\mathrm{R}\cdot\vect{b})
     |\psi\rangle
    +\langle\psi|
     (\gvect{\sigma}_\mathrm{L}\cdot\vect{a'})(\gvect{\sigma}_\mathrm{R}\cdot\vect{b'})
     |\psi\rangle
     \Big|,
 \end{split}
\end{equation}
or again
\begin{equation}\label{bof2}
     S^\psi_{\text{strong}}=
     \Big|
     \langle\psi|
     (\gvect{\sigma}_\mathrm{L}\cdot\vect{a})(\gvect{\sigma}_\mathrm{R}\cdot\vect{b})
    -(\gvect{\sigma}_\mathrm{L}\cdot\vect{a})(\gvect{\sigma}_\mathrm{R}\cdot\vect{b'})
    +(\gvect{\sigma}_\mathrm{L}\cdot\vect{a'})(\gvect{\sigma}_\mathrm{R}\cdot\vect{b})
    +(\gvect{\sigma}_\mathrm{L}\cdot\vect{a'})(\gvect{\sigma}_\mathrm{R}\cdot\vect{b'})
    |\psi\rangle
      \Big|.
\end{equation}
Note, however, that the four spin correlation observables in this
equation are \emph{non commuting observables} (this can be shown
by calculating the commutator of
$(\gvect{\sigma}_\mathrm{L}\cdot\vect{u})(\gvect{\sigma}_\mathrm{R}\cdot\vect{v})$
and
$(\gvect{\sigma}_\mathrm{L}\cdot\vect{u})(\gvect{\sigma}_\mathrm{R}\cdot\vect{v'})$
with $\vect{v}\neq\vect{v'}$), so that the meaning of their
combination must be questioned. The problem is that it is actually
impossible to find an eigenvector for this combination of
observables. Indeed, this linear combination of observables is,
after factorisation, an operator of the form
$\hat{A}\otimes\hat{B}+\hat{C}\otimes\hat{D}$ with
$[\hat{A},\hat{C}]\neq 0$ and $[\hat{B},\hat{D}]\neq 0$. In the
Hilbert space $\mathcal{H}$, an hypothetical eigenvector
$|\phi\rangle\otimes|\chi\rangle$ (with $\alpha$ being its
eigenvalue) of this operator should satisfy
\begin{equation}\label{vectpropre}
    \big[ \hat{A}\otimes\hat{B}+\hat{C}\otimes\hat{D} \big]
    |\phi\rangle\otimes|\chi\rangle= \alpha
     (|\phi\rangle\otimes|\chi\rangle),
\end{equation}
that is,
\begin{equation}
     \hat{A}|\phi\rangle\otimes\hat{B}|\chi\rangle
    +\hat{C}|\phi\rangle\otimes\hat{D}|\chi\rangle
    = \alpha (|\phi\rangle\otimes|\chi\rangle).
\end{equation}
This equation can have solutions only if its left hand side can be
factored; that is, either $\hat{A}|\phi\rangle$ and
$\hat{C}|\phi\rangle$, or $\hat{B}|\chi\rangle$ and
$\hat{D}|\chi\rangle$ must be colinear vectors. This, however, can
never happen because both $[\hat{A},\hat{C}]\neq 0$ and
$[\hat{B},\hat{D}]\neq 0$. Hence, Eq.  (\ref{vectpropre}) has no
solution, and the linear combination of observables in Eq.
(\ref{bof2}) has no eigenvector: \emph{it is not an observable},
and thus it can't be given physical meaning. Therefore,
$S^\psi_{\text{strong}}$ is meaningless and is not a proper
equation to use in order to make physical predictions.

Of course, this does not imply that Quantum Mechanics cannot
provide any meaning at all for the CHSH function; it implies only
that this meaning cannot be strongly objective. Indeed, according
to Von Neumann \cite{JVN1}, any linear combination of expectation
values of different observables $\hat{R}$, $\hat{S},\ldots$ is
meaningful in Quantum Mechanics:
\begin{equation}\label{vn1}
    \langle \hat{R}+\hat{S}+\ldots\rangle_\phi=
     \langle \hat{R} \rangle_\phi
    +\langle \hat{S}\rangle_\phi
    +\ldots
\end{equation}
even if $\hat{R}$, $\hat{S},\ldots$ are non commuting observables.
The explanation is that Quantum Mechanics is only a weakly
objective theory \cite{BDE2,BDE3}, and that expectation values
given by Quantum Mechanics are also weakly objective statements,
that is to say, statements relevant to observations. Hence, when
$\hat{R}$, $\hat{S},\ldots$ are non commuting observables, the
expectation values cannot be simultaneously relevant to the same
set of $N$ systems: each expectation value is necessarily relevant
to a distinct set of $N$ systems (all systems being represented by
the quantum state $|\phi\rangle$). Likewise, the only possible
meaning of Eq.  (\ref{bof}) is therefore weakly objective, not
strongly objective as desired.

Since these expectation values are known with certainty, it is
tempting to consider them as counterfactual entities. However,
conterfactuality requires at least measurement compatibility, that
is, commuting observables. The certainty of a contextual
prediction is not sufficient to make it a counterfactual
prediction; in other words, \emph{weakly objective results known
with certainty are not strongly objective results}. Incidentally,
this is also true in the case of perfect correlations, so that as
a general rule, one may not manipulate weakly objective results as
if they were strongly objective.

The local realistic inequality $S^{\rho}_{\text{strong}}$ cannot
be compared with any strongly objective prediction given by
Quantum Mechanics, so that Bell's Theorem cannot be verified with
a strongly objective interpretation given to the CHSH function,
simply because Quantum Mechanics is not a strongly objective
theory.

This restriction is the first part of a refutation of Bell's
theorem, though maybe not conclusive, since the strength of Bell's
Theorem is mainly its amenability to experimental test. Still,
this was necessary, for now that a strongly objective
interpretation is precluded, there is no choice but to rely on the
weakly objective interpretation in order to compare
hidden-variables theories and Quantum Mechanics.

In the next section, a simple method will be provided in order to
obtain a unique and meaningful quantum prediction for the CHSH
function within weak objectivity.

\section{The Weakly objective interpretation : contextual measurements on $4$
distinct sets of $N$ particle pairs}
\subsection{A Quantum mechanical prediction within the weakly objective
interpretation}\label{qmobj2}

It was shown in Section \ref{CHSH} that strong objectivity and
weak objectivity pertain to different physical systems. This
difference should therefore appear in the relevant equations.
Indeed, the correlation expressed in Eq.  (\ref{qcor2}) is
relevant to spin measurements performed on particles that once
constituted a single parent particle. Yet, two particles issued
from two distinct parents never have interacted with each other,
so that spin measurements performed on such particle pairs can not
be correlated. Hence, if left and right spin measurements are
performed on two distinct sets of $N$ particle pairs, instead of
the same set, there should be no correlation, and this property
should appear in a generalised spin correlation function (i.e.
generalised to the case of spin measurements performed on
different sets of particle pairs).

This can be easily done within a quantum theoretical framework by
means of a distinct EPRB space for each set of $N$ particle pairs.
Let $\mathcal{H}_j$ be the EPRB Hilbert space associated with the
$j$th set of particle pairs. In this Hilbert space, the EPRB
gedanken experiment is represented by the singlet state
$|\psi_j\rangle$ (see Section \ref{EPRB}),
\begin{equation}\label{enfj}
  |\psi_j\rangle=\frac{1}{\sqrt{2}}\Big[|+-\rangle_j-|-+\rangle_j\Big].
\end{equation}
The whole CHSH experiment with the four sets of particle pairs can
be expressed then in terms of a new direct product space
$\mathcal{H}_{1234}\equiv\mathcal{H}_1\otimes\mathcal{H}_2
\otimes\mathcal{H}_3\otimes\mathcal{H}_4$ in which the state
vector is
\begin{equation}\label{vectorpos}
  |\psi_{1234}\rangle=
  |\psi_1\rangle\otimes|\psi_2\rangle\otimes|\psi_3\rangle\otimes|\psi_4\rangle.
\end{equation}
The counterparts of observables in $\mathcal{H}_{1234}$ are
obtained as in Section \ref{observing}. For instance, the
observables pertaining to the right Stern-Gerlach device for the
1st, 2nd, 3rd and 4th set of particle pairs are respectively
\begin{subequations}
    \label{counterp}
    \begin{eqnarray}
  \gvect{\sigma}_{1,\mathrm{R}}\cdot\vect{u}\equiv
  (\gvect{\sigma}_\mathrm{R}\cdot\vect{u})
  \otimes 1\negmedspace\mathrm{l}_2
  \otimes 1\negmedspace\mathrm{l}_3
  \otimes 1\negmedspace\mathrm{l}_4,
  \\
  \gvect{\sigma}_{2,\mathrm{R}}\cdot\vect{u}\equiv
  1\negmedspace\mathrm{l}_1
  \otimes(\gvect{\sigma}_\mathrm{R}\cdot\vect{u})
  \otimes 1\negmedspace\mathrm{l}_3
  \otimes 1\negmedspace\mathrm{l}_4,
  \\
  \gvect{\sigma}_{3,\mathrm{R}}\cdot\vect{u}\equiv
  1\negmedspace\mathrm{l}_1
  \otimes 1\negmedspace\mathrm{l}_2
  \otimes(\gvect{\sigma}_\mathrm{R}\cdot\vect{u})
  \otimes 1\negmedspace\mathrm{l}_4,
  \\
  \gvect{\sigma}_{4,\mathrm{R}}\cdot\vect{u}\equiv
  1\negmedspace\mathrm{l}_1
  \otimes 1\negmedspace\mathrm{l}_2
  \otimes 1\negmedspace\mathrm{l}_3
  \otimes(\gvect{\sigma}_\mathrm{R}\cdot\vect{u}),
    \end{eqnarray}
\end{subequations}
where $1\negmedspace\mathrm{l}_j$ is the identity operator of the
EPRB space $\mathcal{H}_j$. Hence, the expectation value of the
product of two spin observables, the first belonging to the $k$th
set and the second to the $l$th set, is
\begin{equation}\label{expectation2a}
     E^\psi_{kl}(\vect{u},\vect{v})\equiv
     \langle\psi_{1234}|
     (\gvect{\sigma}_{k,L}\cdot\vect{u})(\gvect{\sigma}_{l,R}\cdot\vect{v})
     |\psi_{1234}\rangle,
\end{equation}
and this is the \emph{generalised expectation value of spin
correlation observables} that was sought. The expectation value
for measurements performed on the same set ($k=l$) of particle
pairs is already known, Eq. (\ref{qcor}), and
$E^\psi_{kk}(\vect{u},\vect{v})$ should provide the same result.
Indeed, using Eqs. (\ref{vectorpos}) and (\ref{counterp}) leads to
\begin{align}\label{expectation2b}
    E^\psi_{kk}(\vect{u},\vect{v})&=
    \langle\psi_k|(\gvect{\sigma}_{L}\cdot\vect{u})\cdot(\gvect{\sigma}_{R}\cdot\vect{v})
    |\psi_k\rangle
    \\\nonumber
    &=-\vect{u}\cdot\vect{v},
\end{align}
but when $k\neq l$, the result is quite different:
\begin{align}\label{expectation2c}
     E^\psi_{\begin{subarray}{l} {kl}\\ \nonumber
                    k\neq l
             \end{subarray}}
     (\vect{u},\vect{v})&=
      \langle\psi_k|(\gvect{\sigma}_{L}\cdot\vect{u})|\psi_k\rangle
     \langle\psi_l|(\gvect{\sigma}_{R}\cdot\vect{v})|\psi_l\rangle
     \\
     &=
      \langle\psi_k|\gvect{\sigma}\cdot\vect{u}\otimes1\negmedspace\mathrm{l}_\mathrm{R}
      |\psi_k\rangle
     \langle\psi_l|1\negmedspace\mathrm{l}_\mathrm{L}\otimes\gvect{\sigma}\cdot\vect{v}
      |\psi_l\rangle
     \\\nonumber
     &=0,
\end{align}
in accord with Eq.  (\ref{expspin}). There are indeed no
correlations between two sets of particle pairs, as stipulated in
the beginning of this section.

Now, contrary to what was done in Section \ref{qmobj}, it is
possible to proceed here in full accord with the quantum
mechanical postulates, because the spin correlation observables,
Eqs. (\ref{counterp}), are mutually commuting, so that a linear
combination of these commuting observables is an observable as
well. The CHSH experiment can therefore be described by a new
observable
\begin{equation}\label{Sobserv}
\begin{split}
     \hat{S}_{\text{weak}}\equiv
      (\gvect{\sigma}_{1,L}\cdot\vect{a})&(\gvect{\sigma}_{1,R}\cdot\vect{b})
     -(\gvect{\sigma}_{2,L}\cdot\vect{a})(\gvect{\sigma}_{2,R}\cdot\vect{b'})
     \\
     +&(\gvect{\sigma}_{3,L}\cdot\vect{a'})(\gvect{\sigma}_{3,R}\cdot\vect{b})
     +(\gvect{\sigma}_{4,L}\cdot\vect{a'})(\gvect{\sigma}_{4,R}\cdot\vect{b'}),
\end{split}
\end{equation}
and the quantum prediction for the CHSH function within a weakly
objective interpretation is therefore obtained by calculating the
expectation value of the observable $\hat{S}_{\text{weak}}$ when
the system is in the quantum state $|\psi_{1234}\rangle$ :
\begin{equation}\label{Sq4a}
    S^{\psi}_{\text{weak}}=
    \Big|
    \langle\psi_{1234}|
     \hat{S}_{\text{weak}}
    |\psi_{1234}\rangle
    \Big|,
\end{equation}
which using Eqs.  (\ref{counterp}) and (\ref{expectation2a}) is
\begin{equation}\label{Sq4c}
\begin{split}
     S^{\psi}_{\text{weak}}=
     \Big|
     \langle\psi_1|
     (\gvect{\sigma}_L\cdot\vect{a})&(\gvect{\sigma}_R\cdot\vect{b})
     |\psi_1\rangle
     -\langle\psi_2|
     (\gvect{\sigma}_L\cdot\vect{a'})(\gvect{\sigma}_R\cdot\vect{b})
     |\psi_2\rangle
     \\
     +&\langle\psi_3|
     (\gvect{\sigma}_L\cdot\vect{a})(\gvect{\sigma}_R\cdot\vect{b'})
     |\psi_3\rangle
    +\langle\psi_4|
     (\gvect{\sigma}_L\cdot\vect{a'})(\gvect{\sigma}_R\cdot\vect{b'})
     |\psi_4\rangle
     \Big|,
\end{split}
\end{equation}
that is, using Eq.  (\ref{expectation2b}),
\begin{equation}\label{Sq4e}
     S^{\psi}_{\text{weak}}=\Big|
      E^\psi_{11}(\vect{a},\vect{b})
     -E^\psi_{22}(\vect{a},\vect{b'})
     +E^\psi_{33}(\vect{a'},\vect{b})
     +E^\psi_{44}(\vect{a'},\vect{b'})
     \Big|.
\end{equation}
This equation is not ambiguous (as was Eq. \ref{bof}): it is a
linear combination of expectation values, each relevant to a
distinct set of $N$ particle pairs. This equation is therefore
weakly objective, as requested.

Finally, using Eq.  (\ref{expectation2b}), yields
\begin{equation}\label{Sq4d}
     S^{\psi}_{\text{weak}}=
     \Big|
      \vect{a}\cdot\vect{b}
     -\vect{a}\cdot\vect{b'}
     +\vect{a'}\cdot\vect{b}
     +\vect{a'}\cdot\vect{b'}
     \Big|,
\end{equation}
with a well known maximum equal to
\begin{equation}\label{Sq4f}
     \max(S^{\psi}_{\text{weak}})=2\sqrt{2}.
\end{equation}
This numerical result is indeed the one given in the literature,
the only difference here being the fact that the meaning of this
result is unambiguously weakly objective. Quantum Mechanics, which
is a weakly objective theory \cite{BDE2}, provides a clear answer
to the CHSH function understood as a weakly objective question.

\subsection{A Local realistic inequality within the weakly objective
interpretation}\label{lrweak}

The last step consists in comparing the quantum prediction
$S^{\psi}_{\text{weak}}$ with its local realistic counterpart
$S^{\rho}_{\text{weak}}$. As was stressed in Section \ref{CHSH},
the $j$th set of particle pairs must be characterised by a
distinct set of hidden-variables parameters
$\{\lambda_{j,i}\,;\,i=1,\ldots,N \}$. Hence, to the generalised
expectation value of the spin correlation observable Eq.
(\ref{expectation2a}) corresponds the \emph{generalised mean value
of joint spin measurements}:
\begin{equation}
  M^\rho_{kl}(\vect{u},\vect{v})\equiv
  \frac{1}{N}\sum_{i=1}^{N}
  A(\vect{u},\lambda_{k,i})B(\vect{v},\lambda_{l,i}),
\end{equation}
which is \emph{a priori} capable of reproducing not only the $k=l$
prediction, Eq. (\ref{expectation2b}), but also the $k \neq l$
prediction, Eq. (\ref{expectation2c}). The local realistic CHSH
function with a weakly objective interpretation is therefore
\begin{equation}\label{Srhopos1}
  S^{\rho}_{\text{weak}}=
  \Big|
   M^\rho_{11}(\vect{a},\vect{b})
  -M^\rho_{22}(\vect{a},\vect{b'})
  +M^\rho_{33}(\vect{a'},\vect{b})
  +M^\rho_{44}(\vect{a'},\vect{b'})
  \Big|,
\end{equation}
and that is explicitly
\begin{equation}\label{Srhopos2}
\begin{split}
  S^{\rho}_{\text{weak}}=
  \Big|
  \frac{1}{N}\sum_{i=1}^{N}
  \big[
   A(\vect{a},\lambda_{1,i})&B(\vect{b},\lambda_{1,i})
 -A(\vect{a},\lambda_{2,i})B(\vect{b'},\lambda_{2,i})
  \\
  +&A(\vect{a'},\lambda_{3,i})B(\vect{b},\lambda_{3,i})
  +A(\vect{a'},\lambda_{4,i})B(\vect{b'},\lambda_{4,i})
  \Big]
  \Big|.
\end{split}
\end{equation}

This expression is to be compared with the one pertaining to the
strongly objective interpretation, Eq. (\ref{Srhoobj0}), which
contained terms that could be factored. Here, since each term is
different from the others, no factorisation is possible; i.e.,
\emph{there is no way to derive a Bell inequality}. This is not
the first time this fact has been noticed (see A. Bohm pp. 351,
352 \cite{AB1}), unfortunately, no conclusion was drawn then. Yet,
this fact cannot be ignored, for it has been shown in Section
\ref{qmobj} that Bell's Theorem cannot be demonstrated within a
strongly objective interpretation.

Here, the only local realistic inequality that can be derived is
obtained by considering (as was done with Eq. \ref{rhoobj1}) the
possible numerical values of each term of the summation in Eq.
(\ref{Srhopos2}), that is,
\begin{equation}\label{rhopos1}
\begin{split}
     A(\vect{a},\lambda_{1,i})B(\vect{b},\lambda_{1,i})
    -A(\vect{a},\lambda_{2,i})&B(\vect{b'},\lambda_{2,i})
    +A(\vect{a'},\lambda_{3,i})B(\vect{b},\lambda_{3,i}) \\
    +&A(\vect{a'},\lambda_{4,i})B(\vect{b'},\lambda_{4,i})
    =+4,\ \ +2,\ \ 0,\ \ -2,\ \ -4 \; ,
\end{split}
\end{equation} for which the extrema are +4 and
-4, so that the narrowest local realistic inequality that can be
derived from Eq.  (\ref{Srhopos2}) is nothing but
\begin{equation}\label{Srhopos3}
    S^{\rho}_{\text{weak}}\leq 4.
\end{equation}
This most restrictive local realistic inequality (which can also
be found in Accardi\cite{Accardi}) is not incompatible with the
quantum mechanical prediction, as the maximum of
$S^{\psi}_{\text{weak}}$ is $2\sqrt{2}$. This shows that
experiments intended to test Bell's Theorem were unfortunately not
testing the strongly objective inequality (a Bell inequality, Eq.
\ref{locinobj}), but this weakly objective one, Eq.
(\ref{Srhopos3}), since all experimental tests necessarily are
executed in a weakly objective way, due to the irreducible
incompatibility between spin measurements. As was stressed by Sica
\cite{Sica} and Accardi \cite{Accardi}, a local realistic
inequality is nothing but an arithmetic identity, and inequality
(\ref{Srhopos3}) is definitely too lax to be violated by
experimental tests.

\section{Conclusion}

It was shown that Bell's Theorem cannot be derived, either within
a strongly objective interpretation of the CHSH function, because
Quantum Mechanics gives no strongly objective results for the CHSH
function (see Section \ref{qmobj}), or within a weakly objective
interpretation, because the only derivable local realistic
inequality is never violated, either by Quantum Mechanics or by
experiments \mbox{(see Section \ref{lrweak})}. It was demonstrated
that the discrepancy in Bell's Theorem is due only to a
meaningless comparison between $S^{\rho}_{\text{strong}}\leq 2$
and $S^{\psi}_{\text{weak}}=2\sqrt{2}$, where the former is
relevant to a system with $Nf$ degrees of freedom, whereas the
latter to one with $4Nf$ (see Section \ref{CHSH}). The only
meaningful comparison is between the weakly objective local
realistic inequality $S^{\rho}_{\text{weak}}\leq 4$ and the weakly
objective quantum prediction $S^{\psi}_{\text{weak}}=2\sqrt{2}$,
but these results are not incompatible. Bell's Theorem, therefore,
is refuted.



\begin{thebibliography}{}
\bibitem{JSB2}
J. S. Bell, Physics {\bf1}, 195 (1964).
\bibitem{FS1}
F. Selleri, {\it Le grand débat de la mécanique quantique} (Champs
Flammarion, Paris, 1986).
\bibitem{ASP1}
A. Aspect, Nature {\bf 398}, 189 (1999).
\bibitem{EPR1}
A. Einstein, B. Podolsky, and N. Rosen, Phys. Rev. {\bf47}, 777
(1935).
\bibitem{DB1}
D. Bohm, Phys. Rev. {\bf 85}, 166 (1952).
\bibitem{GHSZ1}
D. Greenberger, M. Horne, A. Shimony and A. Zeilinger, Am. J.
Phys. {\bf 58}, 1131 (1990).
\bibitem{ctdl1}
C. Cohen-Tannoudji, B.Diu, and F. Laloë, {\it Mécanique
Quantique}, vol. I (Hermann, Paris, 1973).
\bibitem{AB1}
A. Bohm, {\it Quantum Mechanics, Foundations and applications}
(Springer-Verlag, New York, 1979).
\bibitem{JSB4}
J. S. Bell, in {\it Proceedings of the international School of
physics 'Enrico Fermi', course} IL: {\it Foundations of quantum
mechanics} (Academic, New York, 1971),  p. 171.
\bibitem{JSB8}
J. S. Bell, Epistemological Letters, p. 2 (July, 1975).
\bibitem{CHSH1}
J. F. Clauser, M. A. Horne, A. Shimony and R. A. Holt, Phys. Rev.
Lett. {\bf 23}, 880 (1969).
\bibitem{BDE2}
B. d'Espagnat, {\it Veiled Reality: An Analysis of Present Day
Quantum Mechanical Concepts}, (Addison-Wesley, 1995).
\bibitem{BDE1}
B. d'Espagnat, http://arXiv/quant-ph/9802046
\bibitem{ADR1}
A. Aspect, J. Dalibard, and G. Roger, Phys. Rev. Lett. {\bf 49},
1804 (1982).
\bibitem{Khrennikov1}
A. Khrennikov, http://arXiv/quant-ph/0006017.
\bibitem{JVN1}
J. von Neumann, {\it Mathematical Foundations of Quantum
Mechanics} (Princeton University Press, 1955).
\bibitem{BDE3}
B. d'Espagnat, {\it Conceptual foundations of Quantum Mechanics},
(W.A. Benjamin, Massachusetts, 1976).
\bibitem{Accardi}
L. Accardi, http://arXiv/quant-ph/0007005.
\bibitem{Sica}
L. Sica, Opt. Commun., {\bf 170}, 55 (1999)
\end{thebibliography}
\end{document}